# Structural Monoclinicity and Its Coupling to Layered Magnetism in Few-Layer CrI$_3$

Xiaoyu Guo[1, *], Wencan Jin[2, *], Zhipeng Ye[3], Gaihua Ye[3], Hongchao Xie[1], Bowen Yang[4], Hyun Ho Kim[5], Shaohua Yan[6], Yang Fu[6], Shangjie Tian[6], Hechang Lei[6], Adam W. Tsen[4], Kai Sun[1], Jia-An Yan[7], Rui He[3, +], & Liuyan Zhao[1, +]

[1] *Department of Physics, University of Michigan, 450 Church Street, Ann Arbor, MI 48109, USA*

[2] *Department of Physics, Auburn University, 380 Duncan Drive, Auburn, AL 36849, USA*

[3] *Department of Electrical and Computer Engineering, Texas Tech University, 910 Boston Ave, Lubbock, TX 79409, USA*

[4] *Institute for Quantum Computing, Department of Physics and Astronomy, and Department of Chemistry, University of Waterloo, ON N2L 3G1, Canada*

[5] *School of Materials Science and Engineering, Kumoh National Institute of Technology, Gumi 39177, Korea*

[6] *Department of Physics and Beijing Key Laboratory of Opto-electronic Functional Materials & Micro-nano Devices, Renmin University of China, 100872 Beijing, China*

[7] *Department of Physics, Astronomy & Geosciences, Towson University, Towson, MD 21252, USA*

[*] author contributed equally

[+] corresponding to: lyzhao@umich.edu (L.Z.); Rui.He@ttu.edu (R.H.)

**Abstract:** Using polarization-resolved Raman spectroscopy, we investigate layer number, temperature, and magnetic field dependence of Raman spectra in one- to four-layer CrI$_3$. Layer-number-dependent Raman spectra show that in the paramagnetic phase a doubly degenerated E$_g$ mode of monolayer CrI$_3$ splits into one A$_g$ and one B$_g$ mode in *N*-layer (*N* > 1) CrI$_3$ due to the monoclinic stacking. Their energy separation increases in thicker samples until an eventual saturation. Temperature-dependent measurements further show that the split modes tend to merge upon cooling but remain separated until 10 K, indicating a failed attempt of the monoclinic-to-rhombohedral structural phase transition that is present in the bulk crystal. Magnetic-field-dependent measurements reveal an additional monoclinic distortion across the magnetic-field-induced layered antiferromagnetism-to-ferromagnetism phase transition. We propose a structural change that consists of both a lateral sliding toward the rhombohedral stacking and a decrease in the interlayer distance to explain our experimental observations.

The physical properties of two-dimensional (2D) van der Waals (vdW) materials can be drastically influenced by their interlayer electronic and magnetic coupling, which is sensitive to many factors such as stacking symmetry,[1-5] interlayer angular twist and vertical spacing,[6-9] as well as external stimuli.[10-15] Recently, the van der Waals magnet $CrI_3$ represents an intriguing case in which the stacking symmetry plays a key role in defining the magnetic ground states.[1, 11, 12, 16] Bulk $CrI_3$ has a monoclinic stacking at room temperature and undergoes a structural phase transition to rhombohedral stacking near $T_s \approx 220$ K,[17] whereas exfoliated few-layer $CrI_3$ retains a monoclinic stacking through all temperatures.[18] Theoretical calculations have shown that the rhombohedral and monoclinic stackings favor ferromagnetic (FM) and antiferromagnetic (AFM) interlayer exchange coupling, respectively.[1] Accordingly, bulk $CrI_3$ exhibits an FM order below the Curie temperature ($T_c$) of ~ 61 K,[17] while few-layer $CrI_3$ develops a layered AFM order below the magnetic onset ($T_M$) of ~ 45 K.[19, 20] Recent magnetic force microscopy[21] and Raman spectroscopy[22] studies further show that in bulk $CrI_3$ the surface layers up to a few tens of layers do exhibit layered AFM order, which is possibly attributed to the persistence of monoclinic stacking in the surface. A careful layer-number-dependent study of stacking symmetry and its coupling with magnetism is thus important for understanding the distinct behaviors between surface layers and the deep bulk of 3D $CrI_3$, as well as those in 2D $CrI_3$, which has however remained missing until now.

The intimate coupling between the layered AFM-to-FM and structural monoclinic-to-rhombohedral phase transitions can be controlled by external stimuli. It has been shown that hydrostatic pressure can switch the layered AFM to FM state *via* tuning the layer stacking in bilayer $CrI_3$.[11, 12] While these studies focus on the role of structure in determining the magnetism, an equally important question, that is, how the structure responds to the change of magnetic states, including both the temperature-driven paramagnetic (PM) to layered AFM and the magnetic-field-induced layered AFM to FM phase transitions, is yet to be studied. The relationship between stacking and magnetism is, in fact, a more generic topic to many 2D magnets than just specific to $CrI_3$. In particular, the correlation between stacking symmetry and magnetic structure has also been reported in another two chromium trihalides, $CrBr_3$[4, 23] and $CrCl_3$.[24] Despite some earlier work in relatively thick $CrI_3$ flakes,[18, 22, 25, 26] a comprehensive and thorough layer-number-dependent investigation of stacking symmetries ranging from mono- to few-layer $CrX_3$ (X = Cl, Br, and I) remains missing; so does the structure response to magnetic phase transitions, including both the temperature-driven and the magnetic-field-induced ones.

In this work, we carry out careful layer-number-, temperature-, and magnetic-field-dependent polarized Raman spectroscopy measurements (see experimental details in the Methods section). In contrast to the existing Raman works [13, 20, 22, 27-29] which are devoted to to the magneto-Raman effect of the $A_g$ phonon mode at ~129 cm$^{-1}$, we focus on a doubly degenerate $E_g$ phonon mode at ~ 107 cm$^{-1}$ [30] (see Fig. 1a) defined in the $D_{3d}$ point group in monolayer (1L) CrI$_3$, which has been demonstrated to be sensitive to interlayer stacking in an angle-resolved polarized Raman spectroscopy study.[11, 12, 18] We start with layer-number dependent measurements in the PM phase. Figure 1b shows the Raman spectra of one- to four-layer (1-4L) CrI$_3$ in the frequency range of 95 - 120 cm$^{-1}$ in both linearly parallel and crossed channels at 80 K (greater than the magnetic onset $T_M$ = 45 K). For 1L CrI$_3$, we observe a doubly degenerate $E_g$ mode at 107 cm$^{-1}$ in both parallel and crossed channels with an equal intensity. In contrast, for 2-4L samples, the $E_g$ mode of 1L CrI$_3$ splits into two modes. The variations of the relative intensity between the two modes among samples can be attributed to the incident linear polarization aligning differently with the in-plane crystal axes of 1-4L CrI$_3$. We can first rule out the possibility of Davydov splitting, for which one would expect a monotonic increase in the number of modes as a function of layer number, as opposed to always two modes for 2-4L CrI$_3$ in Fig. 1b. In particular, in 2L CrI$_3$, which we take as an example here, the lattice structure has inversion symmetry in the PM phase.[20, 27, 31] Under the assumption that there is no degeneracy lift by the interlayer stacking, the Davydov split is expected to lead to two doubly degenerated modes, among which the parity even mode ($E_g$) is Raman active, whereas the parity odd mode ($E_u$) should be Raman silent, resulting in only one mode detected in Raman spectra, which is in contrast to our observation of two modes in 2L CrI$_3$. As a result, we can attribute the origin of split modes only to the monoclinic stacking, which lifts the degeneracy of the $E_g(D_{3d})$ mode in 1L CrI$_3$, resulting in two nondegenerate modes, one $A_g(C_{2h})$ and one $B_g(C_{2h})$, in $N$-layer CrI$_3$ ($N > 1$).

We then fit the spectra to a double-Lorentzian profile in the form of $\sum_{i=1}^{N=2} \frac{A_i\left(\frac{\Gamma_i}{2}\right)^2}{(\omega-\omega_i)^2+\left(\frac{\Gamma_i}{2}\right)^2} + C$, where $\omega_i$ is the central frequency, $\Gamma_i$ is the line width, $A_i$ is the peak intensity, and $C$ is a constant background. Figure 1c shows the frequencies of the fitted modes as a function of layer number $N$. The energy separation between the two split modes grows bigger at increasing $N$ and eventually saturates at $N = 13$, indicating an enhanced interlayer coupling in thicker samples. To quantitatively evaluate the evolution of the monoclinic coupling with the layer number, we propose a minimal model based on an array of coupled harmonic oscillators:

$$H = H_0 + \sum_{<ij>} \left[ a(x_i - x_j)^2 + bx_ix_j + c(y_i - y_j)^2 + dy_iy_j \right]$$

with $H_0 = \frac{1}{2} m \sum_i (\dot{x}_i^2 + \dot{y}_i^2) + \frac{1}{2} k \sum_i (x_i^2 + y_i^2)$.

$H_0$ represents the $E_g$ mode at a frequency of $\omega_0 = \sqrt{\frac{k}{m}}$ within each layer, where $m$ and $k$ are the effective mass and spring constant. $x_i$ and $y_i$ are the magnitude of the $E_g$ mode atomic displacement in the $i^{th}$ layer, and $<ij>$ indicates only the coupling between the nearest neighboring layers. We note that the $a(x_i - x_j)^2$ and $c(y_i - y_j)^2$ terms are typically included in a conventional linear chain model,[27] but they do not lead to the splitting of the two observed Raman modes. To account for the monoclinic stacking, we modify the model by adding the $bx_ix_j$ and $dy_iy_j$ terms, which are allowed by the monoclinic $C_{2h}$ symmetry. Diagonalizing $H$ leads to $2N$ nondegenerate eigenfrequencies in $N$-layer CrI$_3$, and the modes with eigenvectors of $(1,0,1,0\cdots 1,0)$ and $(0,1,0,1\cdots 0,1)$ under the $b \to 0$ and $d \to 0$ limit correspond to the two split modes we observe. Our calculations show that the energy separation is linearly proportional to the difference between $b$ and $d$, i.e., $\Delta\omega = f|b - d|$, and the coefficient $f$ saturates as $N$ increases. As shown in Fig. 1d, by setting $|b - d| = 0.79 \ cm^{-1}$ and all other parameters to be unity, our modified linear chain model can reproduce our experimental data well (see Supplementary Section 1 for detailed calculations).

To pin down the Raman tensors for the two split modes in $N$-layer CrI$_3$ with monoclinic stacking ($N > 1$), we further carry out Raman selection rule measurements on 2L CrI$_3$ under the circular polarization basis. As shown in Fig. 1e, the split $A_g/B_g$ modes observed in the linear polarization channels dominantly appear in the circularly crossed channels (LR and RL), while the signal in the circularly parallel channels (LL and RR) is negligible, where L/R(R/L) stands for the polarization channel with the left/right (right/left)-handed circular polarization for the incident (scattered) light. Only one broad single peak is observed in the LR/RL channel because the $A_g$ and $B_g$ modes are spectrally too close to be resolved when they appear in the same channel. These observations allow us to write the Raman tensors as

$$E_g = \begin{pmatrix} m & g \\ g & -m \end{pmatrix}, \quad A_g = \begin{pmatrix} m & 0 \\ 0 & -m \end{pmatrix}, \quad B_g = \begin{pmatrix} 0 & g \\ g & 0 \end{pmatrix},$$

so that the joint tensor $A_g+B_g = \begin{pmatrix} m & g \\ g & -m \end{pmatrix}$ satisfies the selection rules shown in Fig. 1e. It is worth noting that the Raman tensor of the $A_g$ mode has only one independent element in comparison to the generic form of the $A_g$ mode in the $C_{2h}$ point group $\begin{pmatrix} m & 0 \\ 0 & n \end{pmatrix}$, which corresponds to a specific configuration of monoclinic stacking that constrains $n = -m$.

Having established the characteristics of the monoclinic stacking in $N$-layer CrI$_3$, we are ready to investigate this structural response of monoclinic stacking to magnetic phase transitions. We use 2L CrI$_3$ as a prototypical example in the following studies unless otherwise specified. We first carry out temperature-dependent measurements in linear polarization channels over a wide range from room temperature to 10 K. The crystal axis is carefully aligned so that the split modes appear separately in the parallel and crossed channels. As shown in Fig. 2a, the two split modes both exhibit a significant blue shift upon cooling. It is found that the energy separation between the two modes decreases in this process (see Fig. 2b). Unlike bulk CrI$_3$, where the $A_g$ and $B_g$ modes eventually merge across the monoclinic-to-rhombohedral structural phase transition,[18] in 2L CrI$_3$, despite a tendency toward rhombohedral stacking, the finite separation between the split modes, although small, demonstrates that the monoclinic stacking persists down to 10 K. Moreover, the absence of any abrupt change in the energy separation across $T_{M, c, s}$ indicates that the structural evolution in this process follows a gradual manner. However, as shown in Fig. 2c, the Raman intensities of the split modes suddenly decrease below $T_M$, which originates from the spin-phonon-coupled scattering mechanism.[27]

We then carry out magnetic-field-dependent measurements in circular polarization channels at 10 K. It is known that 2L CrI$_3$ undergoes a layered AFM-to-FM phase transition at a critical field ($B_c$) around $\pm 0.6$ T.[19] Figure 3a shows the Raman spectra acquired in LL and RR channels in the frequency range of 95 - 120 cm$^{-1}$ at 0 and $\pm 2$ T. At 0 T, in contrast to the case of the PM phase (Fig. 1e), the signal of the joint $A_g/B_g$ mode at 107 cm$^{-1}$ (circle) is enhanced in the layered

AFM state. Moreover, two additional modes emerge at ~104 cm$^{-1}$ (square) and ~115 cm$^{-1}$ (diamond), and they obey the same selection rule as the 107 cm$^{-1}$ mode, indicating these two modes are also derived from the joint A$_g$/B$_g$ mode. These observations allow us to write their Raman tensor as

$$A_g + B_g(AFM) = \begin{pmatrix} m & g + h\mathbb{i} \\ g - h\mathbb{i} & -m \end{pmatrix},$$

where the antisymmetric terms break the time-reversal symmetry and account for the layered AFM-assisted phonon scattering.[27] By applying an out-of-plane magnetic field ($B_\perp$) of ±2 T, a Raman circular dichroism emerges for all these three modes, with opposite relative strengths in the LL/RR channels for opposite magnetic field directions. Shown in Fig. 3b is the fitted Raman intensity of the three modes as a function of $B_\perp$. Clearly, the emergence of the Raman circular dichroism happens at the critical magnetic field $B_c$ = ±0.6 T. Such deviation of the selection rule in the magnetic-field-induced FM phase can be attributed to the change in the Raman tensor:

$$A_g + B_g(FM) = \begin{pmatrix} m' & g' + h'\mathbb{i} \\ g' - h'\mathbb{i} & n' \end{pmatrix}.$$

The key finding here is that the constraint in the diagonal tensor elements (*i.e.*, $n' = -m'$) is lifted, and one more independent tensor element is introduced. As a result, the structural contribution in the $A_g + B_g(FM)$ Raman tensor, *i.e.*, the real part, matches the generic form of the monoclinic C$_{2h}$ case, and thus an additional monoclinic structural distortion is detected across this magnetic-field-induced layered AFM-to-FM phase transition.

We then look into the energy separation between the two split modes in the linearly polarized channels to further evaluate this monoclinic distortion across $B_c$. Figure 4a shows the Raman spectra in linear polarization channels at selected magnetic fields (a different sample from that in Fig. 2). The A$_g$/B$_g$ modes can be well resolved in both the linearly parallel and crossed channels at 0 T, while they evolve to nearly merge at 1 T and above. Figures 4b and 4c show the fitted frequencies of the two split modes and their separation as a function of $B_\perp$, respectively. A step-like reduction in the energy separation across $B_c$ is captured, consistent with the first-order nature of the field-induced layered AFM-to-FM transition. We note that the decrease in the doublet frequency separation has also been observed in the temperature-dependent data in Fig. 2b, where upon cooling, a tendency of the monoclinic-to-rhombohedral transition is proposed and a reduction in the interlayer distance was reported previously.[17]

On the basis of our observations above, we present the following interpretation for the response of the monoclinic structure to the magnetic phase transitions in $N$-layer $CrI_3$. To address the suppression of the monoclinic-to-rhombohedral phase transition in $N$-layer $CrI_3$, we carry out density functional theory (DFT) calculations of the total elastic energy as a function of lateral shift between the two layers with respect to the monoclinic stacking. Our results show that there is an energy barrier of ~ 40 meV between the local minimum of monoclinic stacking and the global minimum rhombohedral stacking, which is comparable with the value in the previous report.[1] A plausible explanation is that $N$-layer $CrI_3$ is trapped in the monoclinic local minimum at room temperature, and the thermal energy is not sufficient to overcome this energy barrier. Despite the absence of the monoclinic-to-rhombohedral structural phase transition in $N$-layer $CrI_3$, the lattice structure still undergoes a lateral sliding toward the rhombohedral structure, as evidenced by the decrease in the energy separation of the split modes upon cooling (Fig. 2). Meanwhile, as reported in bulk $CrI_3$,[17] we expect the interlayer distance of few-layer $CrI_3$ also decreases due to thermal contraction. Therefore, we propose that the structural change upon cooling consists of two parts: a lateral sliding toward the rhombohedral stacking and a decrease in the interlayer distance. To reconcile with our modified linear chain model, we comment that first, the values of $b$ and $d$, which quantify monoclinicity, increase, since the reduced interlayer distance enhances the interlayer coupling. Second, the difference between $b$ and $d$, which is correlated to the energy separation between the two modes, decreases, as the adjacent layers slide toward the rhombohedral stacking. This is the scenario that happens across the magnetic-field-induced layered AFM to FM transition at $B_c$, where the reduction in the interlayer distance is responsible for the enhanced interlayer interaction that leads to the emergent Raman circular dichroism and the lateral sliding toward rhombohedral stacking accounts for the decrease of mode frequency separation across $B_c$. Compared to the smooth merging of the two modes upon cooling (Fig. 2b), the adjacent layers experience a more abrupt sliding across $B_c$ that is shown by the sharp step-like drop in the energy separation between the two modes (Fig. 4c).

In summary, we have carried out a comprehensive investigation of the monoclinic stacking of $N$-layer $CrI_3$ and tracked its response to two magnetic phase transitions, the spontaneous thermal one and the magnetic-field-induced spin-flip one in 2L $CrI_3$. A clear monoclinic-stacking-induced degeneracy lift from $E_g$ in monolayer to $A_g+B_g$ in $N$-layer ($N$ = 2, 3, 4, …) $CrI_3$ is demonstrated in the layer-number-dependent results. On one hand, such a monoclinic structure in $N$-layer CrI$_3$ exhibits a failed attempt to transit into the rhombohedral structure upon cooling. On the other hand, this monoclinic structure undergoes a further monoclinic distortion across the magnetic-field-induced layered AFM-to-FM transition. Our results show that the crystal structure also responds intimately to the magnetic phase transition in $N$-layer CrI$_3$, complementary to the reported magnetic phase transition following a change in the stacking symmetry,[11, 12] establishing the presence of magnetoelastic coupling in $N$-layer CrI$_3$ and suggesting a two-way control between structure and magnetism, *i.e.*, changing structure (magnetism) *via* controlling magnetism (structure).

**Methods/Experimental**

**Sample Fabrication.** CrI$_3$ single crystals were grown by the chemical vapor transport method, as detailed in ref. [20, 22, 27, 32]. The CrI$_3$ samples were exfoliated in a nitrogen-filled glovebox. Using a polymer-stamping transfer technique inside the glovebox, we sandwiched CrI$_3$ flakes between two few-layer hBN flakes and transferred them onto SiO$_2$/Si substrates for Raman spectroscopy measurements.

**Raman Spectroscopy.** Micro-Raman spectroscopy measurements were carried out using a 633nm excitation laser. The incident beam was focused by a 40× objective down to ~3 μm in diameter at the sample site, and the power was kept at ~80 μW. The scattered light was collected by the objective in a backscattering geometry, then dispersed by a Horiba LabRAM HR Evolution Raman spectrometer, and finally detected by a thermoelectric-cooled CCD camera. A closed-cycle helium cryostat is interfaced with the Raman system for the temperature-dependent measurements. All thermal cycles were performed at a base pressure that is lower than $7 \times 10^{-7}$ mbar. In addition, a cryogen-free magnet is integrated with the low-temperature cryostat for the magnetic-field-dependent measurements. In this experiment, the magnetic field was applied along the out-of-plane direction and covered a range of 0 to 2.2 T.

**Density Functional Theory Calculations.**

Density-functional theory calculations of monoclinic few-layer CrI$_3$ were performed using the projected augmented-wave (PAW) method as implemented in the Vienna *Ab Initio* Simulation

Package (VASP).[33, 34] In all calculations, we adopted the Perdew–Burke–Ernzerhof (PBE) exchange-correlation functionals, with valence electron configurations as $3d^54s^1$ and $5s^25p^5$ for Cr and I, respectively. van der Waals corrections with optB86b-vdW flavor[35, 36] have been applied to account for the van der Waals interlayer interactions. The Brillouin zone was sampled by a 9 × 9 ×1 Monkhorst-Pack k-point grid mesh, and a 500 eV plane-wave cutoff energy was used. Relaxations were performed until the Hellmann-Feynman force on each atom became smaller than 0.002 eV/Å and the total energy was converged to be within $10^{-6}$ eV. Following intralayer ferromagnetic ordering, magnetic moments for all Cr atoms in the same layer were initialized in the same direction.

**Acknowledgements**

L. Zhao acknowledges the support by NSF CAREER Grant No. DMR-1749774 and AFOSR YIP Grant No. FA9550-21-1-0065. R. He acknowledges the support by NSF CAREER Grant No. DMR-1760668. A. W. Tsen acknowledges support from the US Army Research Office (W911NF-19-10267), Ontario Early Researcher Award (ER17-13-199), and the National Science and Engineering Research Council of Canada (RGPIN-2017-03815). This research was undertaken thanks in part to funding from the Canada First Research Excellence Fund. K. Sun acknowledges the support by NSF Grant No. NSF-EFMA-1741618. H. Lei acknowledges support by the National Key R&D Program of China (Grant No. 2018YFE0202600, 2016YFA0300504), the National Natural Science Foundation of China (No. 11774423, and 11822412), the Beijing Natural Science Foundation (Grant No. Z200005), and the Fundamental Research Funds for the Central Universities and Research Funds of Renmin University of China (RUC) (Grant No. 18XNLG14 and 19XNLG17, 20XNH062). J.A.Yan used the Extreme Science and Engineering Discovery Environment (XSEDE) Comet at the SDSC through allocation TG-DMR160101.

**References**

1. Sivadas, N.; Okamoto, S.; Xu, X.; Fennie, C. J.; Xiao, D., Stacking-Dependent Magnetism in Bilayer CrI$_3$. *Nano Lett* **2018,** *18*, 7658-7664.
2. Bao, W.; Jing, L.; Velasco, J.; Lee, Y.; Liu, G.; Tran, D.; Standley, B.; Aykol, M.; Cronin, S. B.; Smirnov, D.; Koshino, M.; McCann, E.; Bockrath, M.; Lau, C. N., Stacking-Dependent Band Gap and Quantum Transport in Trilayer Graphene. *Nature Physics* **2011,** *7*, 948-952.
3. Lui, C. H.; Li, Z.; Mak, K. F.; Cappelluti, E.; Heinz, T. F., Observation of an Electrically Tunable Band Gap in Trilayer Graphene. *Nature Physics* **2011,** *7*, 944-947.
4. Chen, W.; Sun, Z.; Wang, Z.; Gu, L.; Xu, X.; Wu, S.; Gao, C., Direct Observation of van der Waals Stacking-Dependent Interlayer Magnetism. *Science* **2019,** *366*, 983-987.
5. Shan, Y.; Li, Y.; Huang, D.; Tong, Q.; Yao, W.; Liu, W.-T.; Wu, S., Stacking Symmetry Governed Second Harmonic Generation in Graphene Trilayers. *Sci Adv* **2018,** *4*, eaat0074.
6. Yeh, P.-C.; Jin, W.; Zaki, N.; Kunstmann, J.; Chenet, D.; Arefe, G.; Sadowski, J. T.; Dadap, J. I.; Sutter, P.; Hone, J.; Osgood, R. M., Direct Measurement of the Tunable Electronic Structure of Bilayer MoS$_2$ by Interlayer Twist. *Nano Lett* **2016,** *16*, 953-959.
7. Andrei, E. Y.; Efetov, D. K.; Jarillo-Herrero, P.; MacDonald, A. H.; Mak, K. F.; Senthil, T.; Tutuc, E.; Yazdani, A.; Young, A. F., The Marvels of Moire Materials. *Nat Rev Mater* **2021,** *6*, 201-206.
8. van der Zande, A. M.; Kunstrnann, J.; Chernikov, A.; Chenet, D. A.; You, Y.; Zhang, X.; Huang, P. Y.; Berkelbach, T. C.; Wang, L.; Zhang, F.; Hybertsen, M. S.; Muller, D. A.; Reichman, D. R.; Heinz, T. F.; Hone, J. C., Tailoring the Electronic Structure in Bilayer Molybdenum Disulfide *via* Interlayer Twist. *Nano Lett* **2014,** *14*, 3869-3875.
9. Cao, Y.; Fatemi, V.; Fang, S.; Watanabe, K.; Taniguchi, T.; Kaxiras, E.; Jarillo-Herrero, P., Unconventional Superconductivity in Magic-Angle Graphene Superlattices. *Nature* **2018,** *556*, 43-50.
10. Jiang, S.; Shan, J.; Mak, K. F., Electric-Field Switching of Two-Dimensional van der Waals Magnets. *Nature Materials* **2018,** *17*, 406-410.
11. Li, T.; Jiang, S.; Sivadas, N.; Wang, Z.; Xu, Y.; Weber, D.; Goldberger, J. E.; Watanabe, K.; Taniguchi, T.; Fennie, C. J.; Mak, K. F.; Shan, J., Pressure-Controlled Interlayer Magnetism in Atomically Thin CrI$_3$. *Nature materials* **2019,** *18*, 1303.
12. Song, T.; Fei, Z.; Yankowitz, M.; Lin, Z.; Jiang, Q.; Hwangbo, K.; Zhang, Q.; Sun, B.; Taniguchi, T.; Watanabe, K.; McGuire, M. A.; Graf, D.; Cao, T.; Chu, J.-H.; Cobden, D. H.; Dean, C. R.; Xiao, D.; Xu, X., Switching 2D Magnetic States *via* Pressure Tuning of Layer Stacking. *Nature Materials* **2019,** *18*, 1298-1302.
13. Huang, B.; Cenker, J.; Zhang, X.; Ray, E. L.; Song, T.; Taniguchi, T.; Watanabe, K.; McGuire, M. A.; Xiao, D.; Xu, X., Tuning Inelastic Light Scattering *via* Symmetry Control in the Two-Dimensional Magnet CrI$_3$. *Nature nanotechnology* **2020,** *15*, 212-216.
14. Huang, B.; Clark, G.; Klein, D. R.; MacNeill, D.; Navarro-Moratalla, E.; Seyler, K. L.; Wilson, N.; McGuire, M. A.; Cobden, D. H.; Xiao, D.; Yao, W.; Jarillo-Herrero, P.; Xu, X., Electrical Control of 2D Magnetism in Bilayer CrI$_3$. *Nature Nanotechnology* **2018,** *13*, 544-548.
15. Jiang, S.; Li, L.; Wang, Z.; Mak, K. F.; Shan, J., Controlling Magnetism in 2D CrI$_3$ by Electrostatic Doping. *Nature Nanotechnology* **2018,** *13*, 549-553.
16. Jiang, P.; Wang, C.; Chen, D.; Zhong, Z.; Yuan, Z.; Lu, Z.-Y.; Ji, W., Stacking Tunable Interlayer Magnetism in Bilayer CrI$_3$. *Physical Review B* **2019,** *99*, 144401.

17. McGuire, M. A.; Dixit, H.; Cooper, V. R.; Sales, B. C., Coupling of Crystal Structure and Magnetism in the Layered, Ferromagnetic Insulator CrI$_3$. *Chemistry of materials* **2015,** *27*, 612-620.
18. Ubrig, N.; Wang, Z.; Teyssier, J.; Taniguchi, T.; Watanabe, K.; Giannini, E.; Morpurgo, A. F.; Gibertini, M., Low-Temperature Monoclinic Layer Stacking in Atomically Thin CrI$_3$ Crystals. *2D Materials* **2020,** *7*, 015007.
19. Huang, B.; Clark, G.; Navarro-Moratalla, E.; Klein, D. R.; Cheng, R.; Seyler, K. L.; Zhong, D.; Schmidgall, E.; McGuire, M. A.; Cobden, D. H.; Yao, W.; Xiao, D.; Jarillo-Herrero, P.; Xu, X., Layer-Dependent Ferromagnetism in a van der Waals Crystal Down to the Monolayer Limit. *Nature* **2017,** *546*, 270.
20. Jin, W.; Kim, H. H.; Ye, Z.; Li, S.; Rezaie, P.; Diaz, F.; Siddiq, S.; Wauer, E.; Yang, B.; Li, C.; Tian, S.; Sun, K.; Lei, H.; Tsen, A. W.; Zhao, L.; He, R., Raman Fingerprint of Two Terahertz Spin Wave Branches in a Two-Dimensional Honeycomb Ising Ferromagnet. *Nature Communications* **2018,** *9*, 5122.
21. Niu, B.; Su, T.; Francisco, B. A.; Ghosh, S.; Kargar, F.; Huang, X.; Lohmann, M.; Li, J.; Xu, Y.; Taniguchi, T.; Watanabe, K.; Wu, D.; Balandin, A.; Shi, J.; Cui, Y.-T., Coexistence of Magnetic Orders in Two-Dimensional Magnet CrI$_3$. *Nano Lett* **2020,** *20*, 553-558.
22. Li, S.; Ye, Z.; Luo, X.; Ye, G.; Kim, H. H.; Yang, B.; Tian, S.; Li, C.; Lei, H.; Tsen, A. W.; Sun, K.; He, R.; Zhao, L., Magnetic-Field-Induced Quantum Phase Transitions in a van der Waals Magnet. *Physical Review X* **2020,** *10*, 011075.
23. Si, J.-S.; Li, H.; He, B.-G.; Cheng, Z.-P.; Zhang, W.-B., Revealing the Underlying Mechanisms of Stacking Order and Interlayer Magnetism in Bilayer CrBr$_3$. *arXiv:2011.02720* **2020**.
24. Ahmad, A. S.; Liang, Y.; Dong, M.; Zhou, X.; Fang, L.; Xia, Y.; Dai, J.; Yan, X.; Yu, X.; Dai, J.; Zhang, G.-j.; Zhang, W.; Zhao, Y.; Wang, S., Pressure-Driven Switching of Magnetism in Layered CrCl$_3$. *Nanoscale* **2020,** *12*, 22935-22944.
25. Djurdjic-Mijin, S.; Solajic, A.; Pesic, J.; Scepanovic, M.; Liu, Y.; Baum, A.; Petrovic, C.; Lazarevic, N.; Popovic, Z. V., Lattice Dynamics and Phase Transition in CrI$_3$ Single Crystals. *Physical Review B* **2018,** *98*, 104307.
26. Shcherbakov, D.; Stepanov, P.; Weber, D.; Wang, Y.; Hu, J.; Zhu, Y.; Watanabe, K.; Taniguchi, T.; Mao, Z.; Windl, W.; Goldberger, J.; Bockrath, M.; Lau, C. N., Raman Spectroscopy, Photocatalytic Degradation, and Stabilization of Atomically Thin Chromium Tri-Iodide. *Nano Lett* **2018,** *18*, 4214-4219.
27. Jin, W.; Ye, Z.; Luo, X.; Yang, B.; Ye, G.; Yin, F.; Kim, H. H.; Rojas, L.; Tian, S.; Fu, Y.; Yan, S.; Lei, H.; Sun, K.; Tsen, A. W.; He, R.; Zhao, L., Tunable Layered-Magnetism–Assisted Magneto-Raman Effect in a Two-Dimensional Magnet CrI$_3$. *Proceedings of the National Academy of Sciences - PNAS* **2020,** *117*, 24664-24669.
28. Zhang, Y.; Wu, X.; Lyu, B.; Wu, M.; Zhao, S.; Chen, J.; Jia, M.; Zhang, C.; Wang, L.; Wang, X.; Chen, Y.; Mei, J.; Taniguchi, T.; Watanabe, K.; Yan, H.; Liu, Q.; Huang, L.; Zhao, Y.; Huang, M., Magnetic Order-Induced Polarization Anomaly of Raman Scattering in 2D Magnet CrI$_3$. *Nano Lett* **2020,** *20*, 729-734.
29. McCreary, A.; Mai, T. T.; Utermohlen, F. G.; Simpson, J. R.; Garrity, K. F.; Feng, X.; Shcherbakov, D.; Zhu, Y.; Hu, J.; Weber, D.; Watanabe, K.; Taniguchi, T.; Goldberger, J. E.; Mao, Z.; Lau, C. N.; Lu, Y.; Trivedi, N.; Aguilar, R. V.; Walker, A. R. H., Distinct Magneto-Raman Signatures of Spin-Flip Phase Transitions in CrI$_3$. *Nature Communications* **2020,** *11*, 3879.

30. Webster, L.; Liang, L.; Yan, J.-A., Distinct Spin–Lattice and Spin–Phonon Interactions in Monolayer Magnetic CrI$_3$. *Physical chemistry chemical physics : PCCP* **2018,** *20*, 23546-23555.

31. Sun, Z.; Yi, Y.; Song, T.; Clark, G.; Huang, B.; Shan, Y.; Wu, S.; Huang, D.; Gao, C.; Chen, Z.; McGuire, M.; Cao, T.; Xiao, D.; Liu, W.-T.; Yao, W.; Xu, X.; Wu, S., Giant Nonreciprocal Second-Harmonic Generation from Antiferromagnetic Bilayer CrI$_3$. *Nature* **2019,** *572*, 497-501.

32. Jin, W.; Kim, H. H.; Ye, Z.; Ye, G.; Rojas, L.; Luo, X.; Yang, B.; Yin, F.; Horng, J. S. H.; Tian, S.; Fu, Y.; Xu, G.; Deng, H.; Lei, H.; Tsen, A. W.; Sun, K.; He, R.; Zhao, L., Observation of the Polaronic Character of Excitons in a Two-Dimensional Semiconducting Magnet CrI$_3$. *Nature Communications* **2020,** *11*, 4780.

33. Kresse, G.; Furthmuller, J., Efficient Iterative Schemes for *ab Initio* Total-Energy Calculations Using a Plane-Wave Basis Set. *Physical Review B* **1996,** *54*, 11169.

34. Kresse, G.; Joubert, D., From Ultrasoft Pseudopotentials to the Projector Augmented-Wave Method. *Physical Review B* **1999,** *59*, 1758.

35. Klimes, J.; Bowler, D. R.; Michaelides, A., Chemical Accuracy for the van der Waals Density Functional. *J Phys-Condens Mat* **2010,** *22*, 022201.

36. Klimes, J.; Bowler, D. R.; Michaelides, A., van der Waals Density Functionals Applied to Solids. *Physical Review B* **2011,** *83*, 195131.

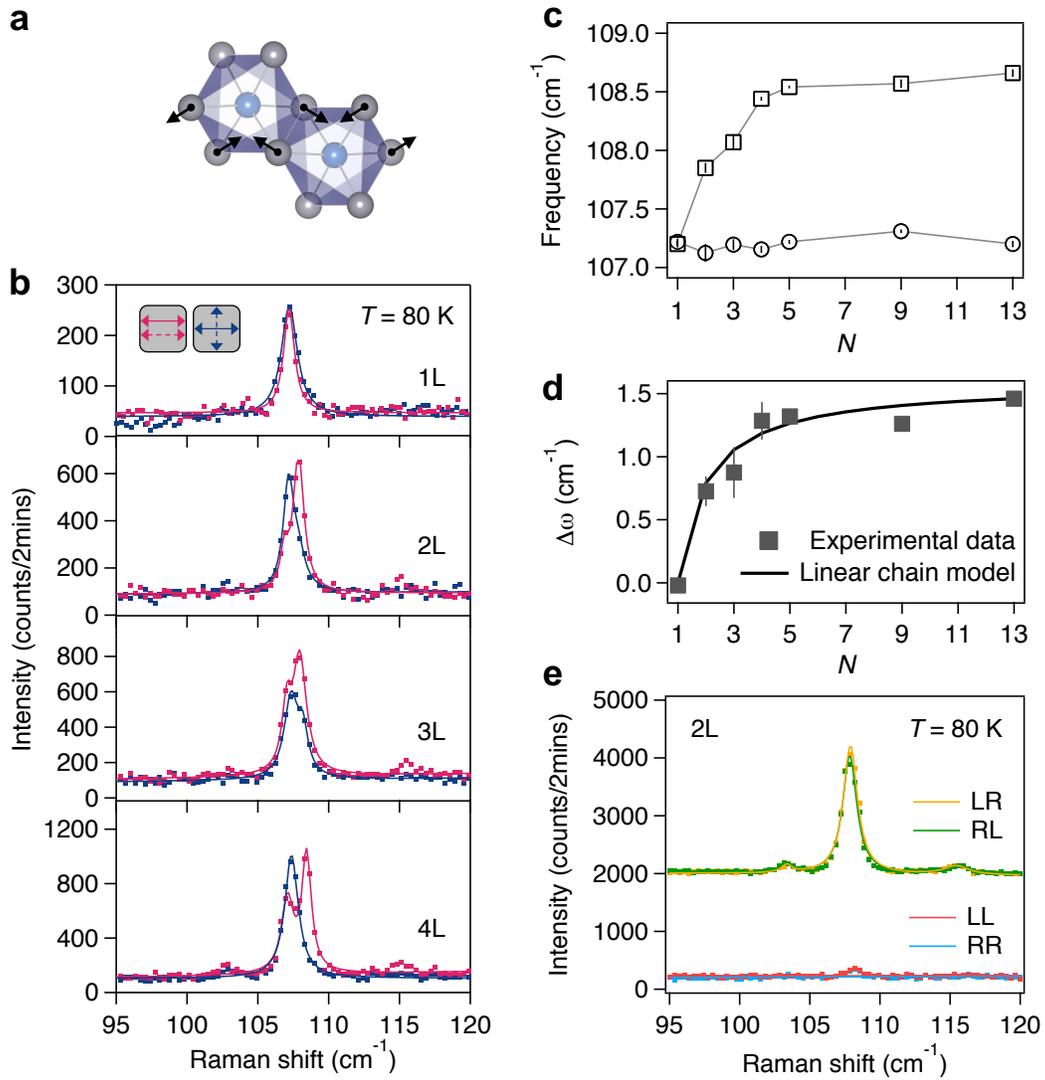

**Fig. 1. a.** Atomic displacement of the $E_g$ mode (107 cm$^{-1}$) in monolayer $CrI_3$. **b.** Raman spectra of 1-4L $CrI_3$ acquired at 80 K in the linearly parallel and crossed polarization channels. The single $E_g$ mode in 1L $CrI_3$ is fitted to a single-Lorentzian profile, and the split $A_g$ and $B_g$ modes in 2-4L $CrI_3$ are fitted to double-Lorentzian profiles. **c.** Plot of the frequencies of the split modes identified in **b** as a function of layer number (*N*). Data from 5L, 9L and 13L samples are included for comparison. **d.** Energy separation ($\Delta\omega$) of the split modes as a function of *N*. Solid squares are experimental data, and the solid curve is a fit to the linear chain model. **e.** Circular polarization selection rules of Raman spectra of 2L $CrI_3$ acquired at 80 K. Error bar stands for one standard error of the fitting parameter.

**Figure 2**

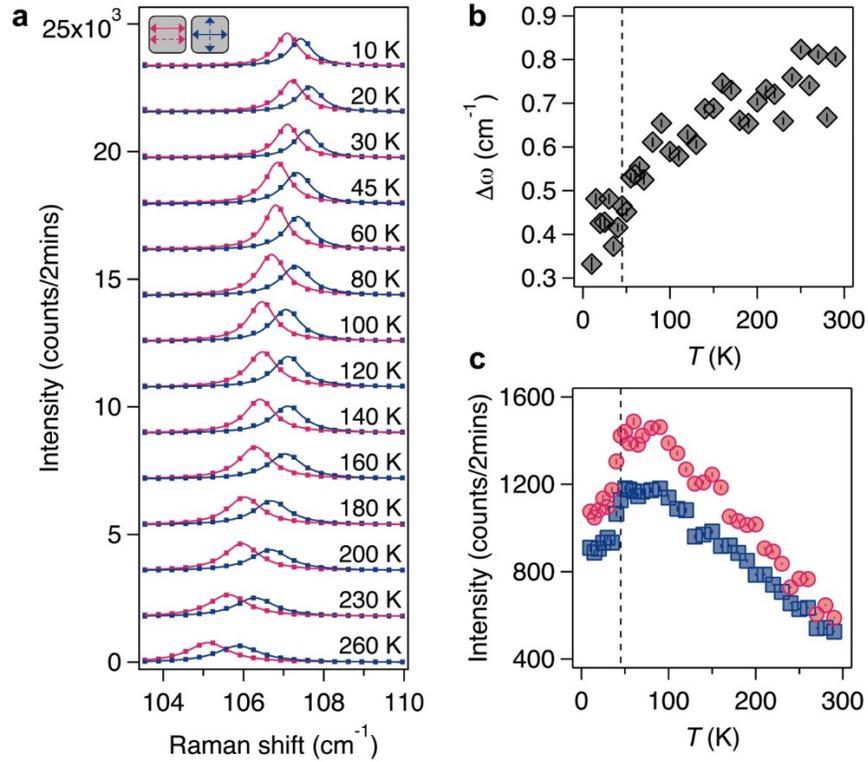

**Fig. 2. a.** Raman spectra (dots are raw data and solid curves are single-Lorentzian fits) of 2L CrI$_3$ acquired at selected temperatures in the linearly parallel and crossed polarization channels. Spectra are vertically offset for clarity. **b.** Plot of the energy separation between the split modes as a function of temperature. **c.** Plots of the intensities of the split modes as a function of temperature. Vertical dashed line in **b** and **c** marks the magnetic transition temperature $T_M$ = 45 K. Error bar stands for one standard error of the fitting parameter.

**Figure 3**

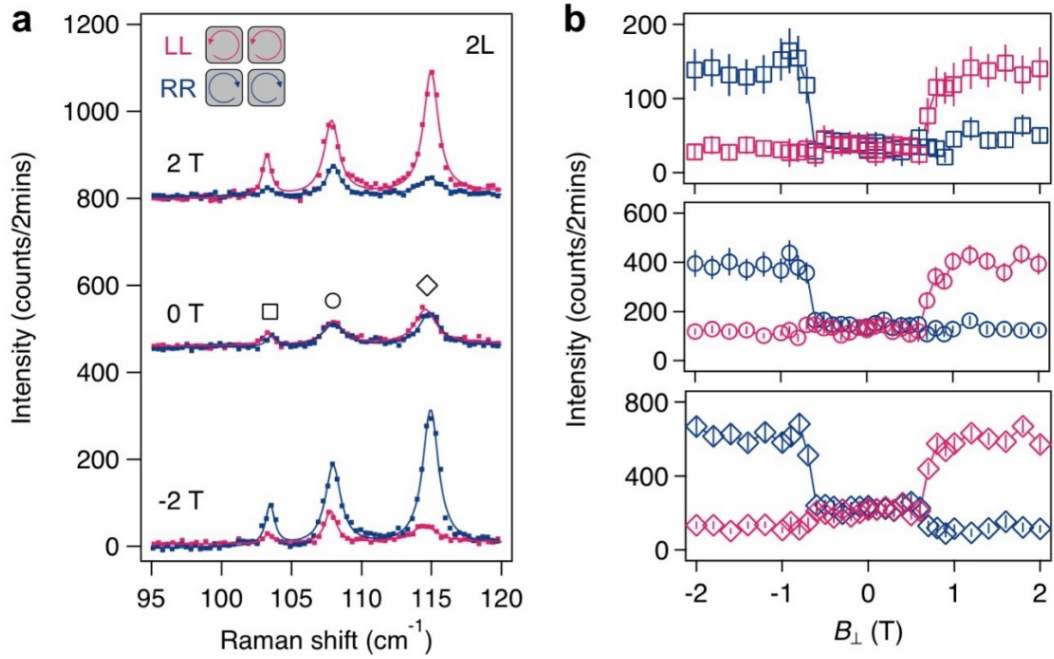

**Fig. 3. a.** Raman spectra of 2L CrI$_3$ acquired at 10 K in the cocircularly polarized LL and RR circular polarization channels with 0 and ±2 T external magnetic fields. **b.** Plot of the intensities of the three Raman modes labeled as square, circle, and diamond in **a** as a function of external magnetic field ($B_\perp$). Error bar stands for one standard error of the fitting parameter.

**Figure 4**

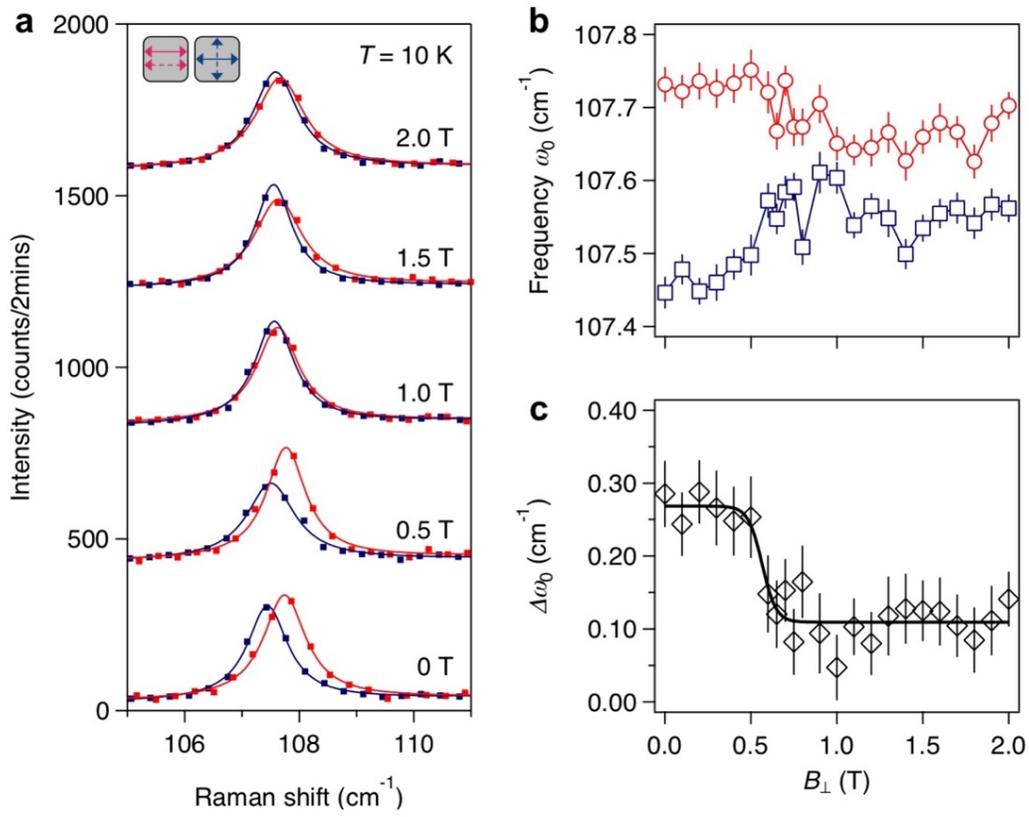

**Fig. 4. a.** Raman spectra of 2L CrI$_3$ acquired at 10 K in the linearly parallel and crossed channels at selected magnetic fields as labeled. **b.** Plot of the fitted frequencies of the two split modes as a function of external magnetic field ($B_\perp$). **c**. Plot of the frequency separation between the two split modes as a function of external magnetic field ($B_\perp$). Error bar stands for one standard error of the fitting parameter.

# Supplementary Information for

**Structural Monoclinicity and Its Coupling to Layered Magnetism in Few-Layer CrI$_3$**

Xiaoyu Guo[1, *], Wencan Jin[2, *], Zhipeng Ye[3], Gaihua Ye[3], Hongchao Xie[1], Bowen Yang[4], Hyun Ho Kim[5], Shaohua Yan[6], Yang Fu[6], Shangjie Tian[6], Hechang Lei[6], Adam W. Tsen[4], Kai Sun[1], Jia-An Yan[7], Rui He[3, +], & Liuyan Zhao[1, +]

[1] *Department of Physics, University of Michigan, 450 Church Street, Ann Arbor, MI 48109, USA*

[2] *Department of Physics, Auburn University, 380 Duncan Drive, Auburn, AL 36849, USA*

[3] *Department of Electrical and Computer Engineering, Texas Tech University, 910 Boston Ave, Lubbock, TX 79409, USA*

[4] *Institute for Quantum Computing, Department of Physics and Astronomy, and Department of Chemistry, University of Waterloo, ON N2L 3G1, Canada*

[5] *School of Materials Science and Engineering, Kumoh National Institute of Technology, Gumi 39177, Korea*

[6] *Department of Physics and Beijing Key Laboratory of Opto-electronic Functional Materials & Micro-nano Devices, Renmin University of China, 100872 Beijing, China*

[7] *Department of Physics, Astronomy & Geosciences, Towson University, Towson, MD 21252, USA*

[*] author contributed equally

[+] corresponding to: lyzhao@umich.edu (L.Z.); Rui.He@ttu.edu (R.H.)

**Table of Contents**

## S1. Linear chain model

As described in the main text, we propose a modified linear chain model in the simplest form of

$$H = H_0 + \sum_{<ij>} \left[ a(x_i - x_j)^2 + b x_i x_j + c(y_i - y_j)^2 + d y_i y_j \right],$$

$$\text{with } H_0 = \frac{1}{2} m \sum_i (\dot{x}_i^2 + \dot{y}_i^2) + \frac{1}{2} k \sum_i (x_i^2 + y_i^2).$$

$H_0$ represents the E$_g$ mode at a frequency of $\omega_0 = \sqrt{\frac{k}{m}}$ within each layer, where $m$ and $k$ are the effective mass and spring constant. $x_i$ and $y_i$ are the magnitude of the E$_g$ mode atomic displacement in the $i^{\text{th}}$ layer, and $<ij>$ indicates only the coupling between the nearest neighboring layers. The two terms in $H_0$ describe the kinetic energy and elastic potential energy of each atomic layer, respectively, and the rest are responsible for the interaction between neighboring layers. The Raman modes we observe in the Raman spectra correspond to the homogeneous A$_g$ and B$_g$ eigenmodes $(x_1, y_1, x_2, y_2 \ldots, x_N, y_N) = (1,0,1,0,\ldots,1,0)$ and $(0,1,0,1,\ldots 0,1)$ under the $b \to 0$ and $d \to 0$ limit, where $N$ is the total layer number. In the Hamiltonian accounting for the interlayer coupling, the $a(x_i - x_j)^2$ and $c(y_i - y_j)^2$ terms represent the energy cost induced by the relative displacement between adjacent layers. In the conventional linear chain model where $b = d = 0$, the frequencies of the homogeneous A$_g$ and B$_g$ modes are solved to be $\omega_1 = \omega_2 = \omega_0$ with no energy splitting.

To reconcile with our experimental finding of the two split modes in $N$-layer CrI$_3$, we introduce the minimally required perturbative corrections $b x_i x_j$ and $d y_i y_j$ ($b \ll a$ and $d \ll c$) which are allowed by the C$_{2h}$ symmetry. We can in principle add more terms that obey the C$_{2h}$ symmetry, but simultaneously will introduce redundant fitting parameters. Here, we decide to choose this minimal modification from the conventional linear chain model and show that it is sufficient in explain our results. Physically, these two terms account for the monoclinic coupling between adjacent layers. By diagonalizing the Hamiltonian, we find that this introduction of monoclinicity leads to the splitting of the A$_g$ and B$_g$ modes. The calculated frequencies $\omega_1$ and $\omega_2$ for layer number $N = 1 \sim 4$ are shown in Table S1. Under the assumption of $b \ll a$ and $d \ll c$, we perform Taylor expansion and find that the energy separation between the two modes is linearly proportional to $|b - d|$. By setting $|b - d| = 0.79 \, cm^{-1}$ and all other parameters to be unity, we plot the energy separation predicted by the linear chain model, together with the experiment results in Fig. 1d in the main text, which shows an excellent agreement between the two.

Table S1. Calculated frequencies of the two homogeneous vibration modes ($\omega_{1,2}$) in 1-4L CrI$_3$ using the modified linear chain model

| N | $\omega_1^2$ | $\omega_2^2$ |
|---|---|---|
| 1 | $\omega_0^2$ | $\omega_0^2$ |
| 2 | $\omega_0^2 + \dfrac{b}{m}$ | $\omega_0^2 + \dfrac{d}{m}$ |
| 3 | $\omega_0^2 + \dfrac{3a - \sqrt{9a^2 - 8ab + 2b^2}}{m}$ | $\omega_0^2 + \dfrac{3c - \sqrt{9c^2 - 8cd + 2d^2}}{m}$ |
| 4 | $\omega_0^2 + \dfrac{4a + b - \sqrt{16a^2 - 16ab + 5b^2}}{m}$ | $\omega_0^2 + \dfrac{4c + d - \sqrt{16c^2 - 16cd + 5d^2}}{m}$ |

When the temperature cools down, the energy separation between the two homogeneous A$_g$ and B$_g$ modes decreases which corresponds to a reduction in the difference between $b$ and $d$. This indicates a lateral sliding between adjacent layers and one can imagine when the layers slide to the rhombohedral stacking, $b$ will equal to $d$ and there will be no splitting. The same change may happen across the layered AFM-to-FM phase transition, though the sharp decrease in the energy separation may indicate a more abrupt and fierce sliding compared to the one in the cool down process.

## S2. Density functional theory calculation of bilayer CrI$_3$

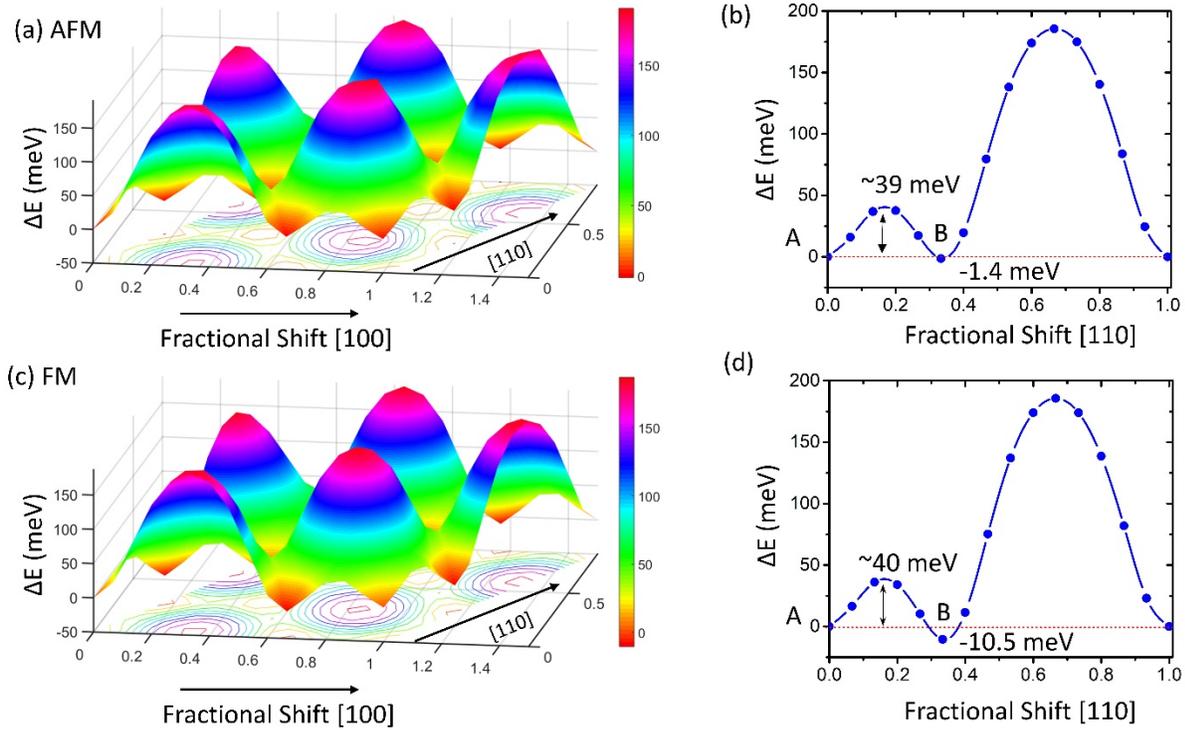

**Fig. S1** The full energy landscape over the lateral fractional shift under (a) AFM and (c) FM interlayer coupling, respectively. (b) and (d) show a line cut of (a) and (c) along the [110] direction. An energy barrier of ~40 meV is observed between the monoclinic stacking (point A) and rhombohedral stacking (point B).

We have investigated the possible structural phase transitions in monoclinic bilayer CrI$_3$ (point group symmetry C$_{2h}$). For both AFM and FM interlayer magnetic orders, we shifted the upper layer with respect to the bottom one with a combined fractional coordinate along the [100] and [110] directions. A total of 16×16 = 256 configurations has been calculated. For each configuration, we fixed the atomic coordinates in the layer plane while allowing the vertical coordinates to get fully optimized. The energy difference (ΔE) relative to the starting monoclinic structure (denoted as A in Fig. S1b and Fig. S1d) has been calculated and the resulted energy landscapes have been shown in Fig. S1a and Fig. S1c for AFM and FM orders, respectively. The corresponding typical energy differences ΔE along the [110] direction have been shown in Fig. S1b and Fig. S1d. As can be seen clearly from the plots, there are energy barriers of about 40 meV to the nearest saddle points (denoted as B in Fig. S1b and Fig. S1d, corresponding to the $R\bar{3}$ rhombohedral stacking order) in both AFM and FM. Such energy barriers might not be easily overcome with the external magnetic field, thus prohibiting a possible structural phase transition.

Note that in the FM order, the energy of B structure will be about -10.5 meV lower than that of A. However, this energy difference becomes much smaller in the AFM order (~1.4 meV).